\documentclass[11pt,a4paper]{article}
\usepackage{jheppub}
\usepackage{graphicx}

\def\hbar{\hspace{0pt}\raisebox{1pt}{$-$} \hspace{-7pt} h}

\def\5{\overline 5}

\newcommand{\ba}{\begin{eqnarray}}
\newcommand{\ea}{\end{eqnarray}}
\newcommand{\no}{\nonumber}
\newcommand{\be}{\begin{equation}}
\newcommand{\ee}{\end{equation}}
\newcommand{\bea}{\begin{eqnarray}}
\newcommand{\eea}{\end{eqnarray}}

\title{Multitrace deformations, Gamow states,\\ and Stability of AdS/CFT}
\date{\today
}
\author{
Luca Vecchi}
\affiliation{Theoretical Division T-2, Los Alamos National Laboratory \\ Los Alamos, NM 87545, USA}
\emailAdd{vecchi@lanl.gov}
\abstract{

We analyze the effect of multitrace deformations in conformal field theories at leading order in a large N approximation. These theories admit a description in terms of weakly coupled gravity duals. We show how the deformations can be mapped into boundary terms of the gravity theory and how to reproduce the RG equations found in field theory. In the case of doubletrace deformations, and for bulk scalars with masses in the range $-d^2/4<m^2<-d^2/4+1$, the deformed theory flows between two fixed points of the renormalization group, manifesting a resonant behavior at the scale characterizing the transition between the two CFT's. On the gravity side the resonance is mapped into an IR non-normalizable mode (Gamow state) whose overlap with the UV region increases as the dual operator approaches the free field limit. We argue that this resonant behavior is a generic property of large N theories in the conformal window, and associate it to a remnant of the Nambu-Goldstone mode of dilatation invariance. We emphasize the role of nonminimal couplings to gravity and establish a stability theorem for scalar/gravity systems with AdS boundary conditions in the presence of arbitrary boundary potentials and nonminimal coupling. 
}

\begin{document}
\maketitle

\section{Introduction}

Multitrace deformations of large $N$ conformal field theories (CFT's) were first studied in the context of the gauge/gravity correspondence in~\cite{first}\cite{Witten}, and soon after in~\cite{preSS}\cite{Mueck}\cite{SS}. These can be described by the following expression
\ba\label{theory}
I_{CFT}+\int d^dx\,W({\cal O}).
\ea
Here, $I_{CFT}$ formally denotes a $d$-dimensional large $N$ CFT, and ${\cal O}$ is a (composite) scalar singlet operator of the CFT with scaling dimension $\Delta$. $W$ is an arbitrary functional density of the scalar operator which will be referred to as a multitrace deformation of the CFT. 

On the gravity side, the $d$-dimensional, large $N$ CFT is described by an $AdS_{d+1}$ geometry~\cite{Maldacena}, whereas ${\cal O}$ is mapped into a scalar mode $\phi$ propagating in the bulk of space-time. The mass $m$ of the bulk scalar, normalized in units of the $AdS$ curvature, is related to the scaling dimension $\Delta$ of the dual operator by
\ba\label{Dim}
\Delta_\pm&=&\frac{d}{2}\pm\nu,\\\no
\nu&=&\sqrt{m^2+\frac{d^2}{4}}.
\ea
The duality between the CFT and the gravity theory completes once appropriate boundary conditions on the scalar $\phi$ are imposed~\cite{AdS/CFT}\cite{Klebanov}. These boundary conditions select which of the quantities $\Delta_\pm$ should be associated to the scaling dimension $\Delta$ of the dual operator ${\cal O}$. In other words, it is the boundary condition that \emph{defines} the dual CFT. 


In the standard AdS/CFT prescription one implements Dirichlet boundary conditions on the bulk fields~\cite{AdS/CFT}\cite{Klebanov}. This way one finds that $\Delta=\Delta_+$. The partition function of the CFT with $\Delta=\Delta_-$ was recovered in~\cite{Klebanov} as a Legendre transform of the former functional. In the standard approach, therefore, the quantizations $\Delta=\Delta_\pm$ are treated differently. 

A study of multitrace deformations of CFT's suggests, on the other hand, that the two inequivalent boundary theories $\Delta=\Delta_\pm$ can be recovered as two limits of the same \emph{deformed} theory. To appreciate this it is useful to refer to calculable, large $N$ field theories defined by the formal structure~(\ref{theory}).

A complete field theory planar analysis of~(\ref{theory}) with $W({\cal O})=\frac{f}{2}{\cal O}^2$ a doubletrace deformation is given in~\cite{PR}\cite{cw}, see also~\cite{Kleba} for a 1-loop analysis. The main conclusions of these studies can be summarized in terms of the beta function of the dimensionless coupling $\bar f=f\Lambda^{2\Delta-d}$
\ba\label{Beta}
\beta_{\bar f}\equiv\Lambda\frac{d\bar f}{d\Lambda}=\bar f^2+(2\Delta-d)\bar f+a,
\ea
and the scaling dimension of the operator ${\cal O}$
\ba
\Delta_{\cal O}=\Delta+\bar f.
\ea
In the above formulas, the parameters $a$ and $\Delta$ are functions of the couplings of the CFT and can be regarded as constants in a leading planar analysis~\cite{PR}. A non-vanishing $a$ term in the beta function~(\ref{Beta}) appears if the doubletrace deformation ${\cal O}^2$ is required as a counterterm, in which case the undeformed theory is not a complete quantum theory, and in particular not a CFT. 
This is for example the case of orbifold projections of ${\cal N}=4$ SYM, see~\cite{Petkou} for an early reference.

The existence of a beta function for $\bar f$ tells us that the deformed theory~(\ref{theory}) is not generally a large $N$ CFT. A look at~(\ref{Beta}) reveals in fact two distinct regimes characterized by the sign of the discriminant of~(\ref{Beta}):
\ba\label{D}
D=\left(\Delta-\frac{d}{2}\right)^2-va.
\ea
If $D\geq0$ the theory admits two fixed points $\bar f_\pm$ of the renormalization group, and the scaling dimensions of the operator ${\cal O}$ at the two fixed points read
\ba\label{Delta}
\Delta_\pm=\frac{d}{2}\pm\sqrt{D}.
\ea
The additional requirement $D<1$ ensures the scalar satisfies the unitarity bound~\cite{Mack} at both fixed points. A consistent description of the theory in the regime $0\leq D<1$ can only be given -- in a leading $1/N$ approximation -- for couplings in the range $\bar f_-<\bar f<\bar f_+$, in which case the dynamics describes a flow between two CFT's. This latter feature was shown to lead to the appearance of a resonant pole in the 2-point functions of the field ${\cal O}$ in~\cite{cw}. For renormalized couplings outside the range $\bar f_-<\bar f<\bar f_+$ the doubletrace deformation is either trivial or it induces a tachyonic instability. 

The model~(\ref{theory}) with $W=\frac{f}{2}{\cal O}^2$ is completely destabilized in the regime $D<0$ in which the two fixed points $\bar f_\pm$ formally move to the complex plane.

Eq.~(\ref{Delta}) represents the field theory expression of the AdS/CFT relation~(\ref{Dim}) between the two inequivalent dimensions $\Delta_\pm$ of a CFT operator ${\cal O}$ and the mass $m$ of the dual bulk scalar $\phi$. It follows that~\cite{PR}
\ba\label{i}
\nu=\sqrt{D}.
\ea
Note that the identification~(\ref{i}) is consistent with the field theory expectation that $D<0$ signals an instability of the CFT. Indeed, it is well known that if the bulk scalar mass is such that $\nu^2<0$ the $AdS_{d+1}$ geometry becomes unstable~\cite{BF}.

From the field theory perspective it thus appears clear that the two inequivalent CFT's associated to the two gravity quantizations $\Delta=\Delta_\pm$ are related by a doubletrace deformation. 
%
This in particular anticipates that there should exist a prescription that allows us to describe the two inequivalent quantizations $\Delta=\Delta_\pm$ in a symmetric way. In addition, the field theory analysis of~\cite{PR}\cite{cw} leaves us with a few interesting questions to be answered: Is there a systematic way to embed multitrace deformations on the gravity theory? Can we reproduce the RG flow found on the field theory side? How does the $a$ term in~(\ref{Beta}) appear on the gravity dual? What is the nature of the resonant mode in the gravity description? Is it a specific property of doubletrace deformed CFT's or a more general feature? The aim of this paper is to address these issues.

A systematic implementation of multitrace deformations on the gravity dual theories was proposed in~\cite{SS}. There it was suggested that CFT deformations can be mapped into appropriate boundary terms on the gravity side. The resulting boundary conditions on the dual scalar $\phi$ agreed with the general recipe proposed in~\cite{Witten}. 

The first part of our work will be hence devoted to a review of the formalism developed in~\cite{AdS/CFT}\cite{Klebanov}\cite{Witten}, and especially in~\cite{SS}. The main difference between our approach and that of~\cite{SS} is that in the latter paper the scaling dimension of the dual operator was assumed to be $\Delta=\Delta_+$ in order to avoid singularities at the conformal boundary. Here we will instead focus on the case $\Delta=\Delta_-$. As we will see, our approach has interesting implications on the analysis of the RG flow of the couplings of multitrace deformations. In fact, the boundary terms on the gravity side will be shown to be in one to one correspondence with multitrace deformations of the \emph{1 Particle Irreducible} (1PI) action of the operator ${\cal O}$ of dimension $\Delta=\Delta_-$, as opposed to being mapped into deformations of the \emph{generating functional} of an operator ${\cal O}$ of dimension $\Delta=\Delta_+$ as in~\cite{SS}. We will find that it is precisely the RG scale dependence hidden in the boundary action that prevents the appearance of singularities on the gravity side. This will allow us to easily identify the RG equations for the boundary couplings. We will eventually use the tools developed in this preliminary part of the paper to derive our main original results. These will be discussed in Sections~\ref{double},~\ref{RES}, and~\ref{stability}.

The paper is organized as follows. In Section~\ref{two} we will review the results of~\cite{AdS/CFT}\cite{Klebanov}\cite{Witten}\cite{SS} relevant to our analysis. Here we will emphasize that the two inequivalent quantizations $\Delta=\Delta_\pm$ can be selected by two distinct boundary conditions on the scalar $\phi$, and specifically that these two quantizations can be treated in a completely symmetric way. Consistently with the general prescription adopted here, the source of the dual operator ${\cal O}$ will be described by a linear term at the boundary of the gravity action. An explicit mapping between arbitrary CFT deformations and boundary terms on the gravity side is presented in Section~\ref{multi}.

Doubletrace deformations, for which the AdS/CFT analysis can be compared to the field theory predictions of~\cite{PR}\cite{cw}, will be studied in some detail in Section~\ref{double}. Generalizing the results of~\cite{Witten} we will see that the leading $1/N$ RG flow~(\ref{Beta}) induced by the doubletrace deformations and the identification $\nu=\sqrt{D}$ proposed in~\cite{PR} can be correctly reproduced by a weakly coupled gravity dual. 
A special role in our discussion is played by nonminimal couplings to gravity which will be shown to be in one to one correspondence with the $a$ term in~(\ref{Beta}).

The weakly coupled gravity description should also be able to reproduce other features of the field theory, including the resonant behavior characterizing doubletrace deformed large N CFT's~\cite{cw}. The resonance has a non-zero width already at leading order in the $1/N$ expansion, and it cannot be found in the physical spectrum of the gravity dual. In Section~\ref{RES} we will show that this state appears as an IR non-normalizable mode on the gravity side, as its composite nature anticipates. These modes were first introduced by Gamow in his study of alpha decay, and are sometimes referred to as Gamow states in the literature. The scalar resonance which appears in these scenarios was associated to a remnant of the Nambu-Goldstone mode of conformal invariance in~\cite{cw}. We will explain how this interpretation is recovered on the gravity side, and argue that the resonant behavior is a generic feature of theories flowing between two CFT's.

The theorem for the classical stability of $AdS_{d+1}$ discussed in~\cite{des2}\cite{des3}\cite{des4} in the presence of a generic (potential) deformation $W$ will be generalized to models with nonminimal couplings in Section~\ref{stability}. In agreement with field theory expectations, the stability of the nonminimally coupled systems will be found to depend on the bulk dynamics as well as on the boundary potential, in sharp contrast with the minimally coupled systems.

\section{A systematic approach}
\label{systematic}

There exist two inequivalent quantizations for a scalar field in $AdS_{d+1}$ with bulk mass in the range ($0<\nu<1$)~\cite{BF}
\ba
m_{BF}^2<m^2<m^2_{BF}+1,
\ea
where $m^2_{BF}=-d^2/4$. The importance of this fact in the context of the AdS/CFT correspondence has been revealed in~\cite{Klebanov}, where it was argued that the two quantizations are associated to the two scaling dimensions $\Delta_\pm$ defined in~(\ref{Dim}).

In this section we build on the results of~\cite{SS} and eventually develop a formalism that allows us to treat the two quantizations $\Delta=\Delta_\pm$ in a symmetric way. A natural generalization of this prescription will then be used to systematically encode the effect of a generic multitrace deformation $W({\cal O})$ along the lines of~\cite{Witten}\cite{SS}.

\subsection{Two possible quantizations}
\label{two}

Let us begin by introducing our notation. We define the $AdS_{d+1}$ metric as:
\ba\label{AdS}
ds^2=\frac{1}{z^2}(\eta_{\mu\nu}dx^\mu dx^\mu-dz^2),
\ea
where for simplicity the curvature length has been normalized to $1$. We will be formulating our gravity theory in the regularized region $z>\epsilon$, where $\epsilon^{-1}$ may be viewed as an UV cutoff of the dual theory. At the end of the computation we will eventually consider the continuum limit $\epsilon\rightarrow0$.

For finite UV cutoff, the quadratic theory for a scalar $\phi$ contains a boundary term. In Euclidean space this is:
\ba\label{boundary}
\int d^dx\sqrt{h}\left[\frac{\lambda'}{2}(\partial\phi)^2+\frac{\Delta_{\cal O}}{2}\phi^2+\tilde J\phi\right]_{z=\epsilon},
\ea
where $h$ is the determinant of the induced metric at the boundary $z=\epsilon$. The boundary kinetic term may be neglected with respect to the mass term as far as $\lambda'(q\epsilon)^2\ll\Delta_{\cal O}$. In the limit $\epsilon\ll1$ and for $d$-dimensional momenta below the UV cutoff, i.e. $q\epsilon<1$, we can safely ignore it and set $\lambda'=0$. This choice is solely dictated by simplicity: the main results presented in this paper can be generalized to the case $\lambda'\neq0$. Under our assumption, the regularized quadratic action for $\phi$ can be written on the $AdS$ background~(\ref{AdS}) as
\ba\label{reg}
I_{reg}&=&\frac{1}{2}\int d^dx\int dz\, z^{1-d}\left[\partial\phi^2+\phi'^2+\frac{m^2}{z^2}\phi^2\right]\\\no
&+&\int_{z=\epsilon} d^dx z^{-d}\left[\frac{\Delta_{\cal O}}{2}\phi^2+\tilde J\phi\right],
\ea
where a prime denotes derivative with respect to $z$. The linear term $\tilde J\phi$ will be identified with the source of the CFT operator dual to $\phi$ below, whereas the quadratic term $\Delta_{\cal O}\phi^2$ with a doubletrace deformation. At this stage this latter term may be seen as a counterterm in the language of~\cite{HolRG}.

From the regularized theory~(\ref{reg}) we derive the equations of motion
\ba\label{EOM}
-\phi''+\frac{d-1}{z}\phi'+\frac{m^2}{z^2}\phi=\partial^2\phi,
\ea
and the boundary conditions
\ba\label{bcon}
z^{-d}\delta\phi\left[z\phi'-\Delta_{\cal O}\phi-\tilde J\right]_{\epsilon}=0,\\\no z^{-d}\delta\phi\left[z\phi'\right]_\infty=0.
\ea
The general solution for $m^2>m_{BF}^2$ behaves for small $z$ as
\ba\label{sol}
\phi&=&\alpha z^{\Delta_-}(1+O(z^2))+\beta z^{\Delta_+}(1+O(z^2)),
\ea
where $\Delta_\pm$ are given in~(\ref{Dim}), whereas $\alpha$ and $\beta$ are generic functions of the transverse coordinates $x^\mu$. The solution for the degenerate case $m^2=m^2_{BF}$ behaves as 
\ba\label{sol'}
\phi=z^{d/2}(\alpha \log z+\beta)(1+O(z^2)),
\ea
while for $m^2<m_{BF}^2$ the exponents $\Delta_\pm$ become complex and the scalar $\phi$ propagates tachyons. In the following we wil assume that $m^2\geq m_{BF}^2$. The solutions with $\alpha=0$ or $\beta=0$ are also referred to in the literature as fast or slow falloff solutions, or even as regular or irregular solutions, respectively.

The standard AdS/CFT prescription~\cite{AdS/CFT}\cite{Klebanov} is formulated with Dirichlet boundary conditions $\delta\phi|_\epsilon=0$. In this case the field $\phi$ is fixed at $z=\epsilon$, and its boundary value $\alpha$ is interpreted as the CFT source $J$. We will call this the $\Delta_+$ quantization, because the operator dual to $\phi$ turns out to have scaling dimension $\Delta_+$. The $\Delta_-$ quantization, in which the dual operator has weight $\Delta_-$, is obtained by identifying $\beta$ with the source of the dual CFT operator. 

We wish to formulate a prescription such that the relations $\alpha=J$ or $\beta=J$ are automatically induced by suitable boundary conditions. We thus require the scalar to satisfy (see~(\ref{bcon}))
 \ba\label{bc}
\left[z\phi'-\Delta_{\cal O}\phi\right]_{\epsilon}=\left[\tilde J\right]_{\epsilon}
\ea
at the UV boundary $z=\epsilon$, whereas the condition in the interior follows from regularity of the solution, as indicated in~(\ref{bcon}). A discussion of the generalized prescription~(\ref{bc}) first appeared in~\cite{Minces1}\cite{Minces2}\cite{SS}.

The advantages of the formulation~(\ref{bc}) are manifold. First, the prescription~(\ref{bc}) allows us to treat the two quantizations $\Delta=\Delta_\pm$ is a symmetric fashion:  the $\Delta_\pm$ quantizations are simply selected by choosing $\Delta_{\cal O}=\Delta_\pm$. Second, a straightforward generalization of~(\ref{bc}) can be used to systematically implement the boundary conditions introduced in~\cite{Witten} to account for the presence of an arbitrary multitrace deformation $W$ (this was first done in~\cite{SS} for $\Delta_{\cal O}=\Delta_+$). Third, the formulation~(\ref{bc}) suggests a simple and intuitive way of deriving the Wilsonian RG flow of the boundary couplings of the multitrace deformation.

The last two points will be clarified in Sections~\ref{multi} and~\ref{RG} respectively. In this subsection we will emphasize that the prescription~(\ref{bc}) implies that the choice of quantization (whether $\Delta_-$ or $\Delta_+$) is mapped into a choice of the boundary mass term $\Delta_{\cal O}$. This is readily understood by observing that, with the help of the asymptotic forms~(\ref{sol}) and~(\ref{sol'}), the boundary condition~(\ref{bc}) with $\Delta_{\cal O}=\Delta_+$ ($\Delta_-$) reads $2\nu\alpha=-\tilde J\epsilon^{-\Delta_-}$ ($2\nu\beta=\tilde J\epsilon^{-\Delta_+}$), which -- up to an irrelevant normalization -- basically represents the identification proposed in~\cite{Klebanov}.  

In order to unambiguously prove the consistency of our approach we now show that the regularized action~(\ref{reg}) evaluated on a solution satisfying~(\ref{bc}) with $\Delta_{\cal O}=\Delta_\pm$ correctly reproduces the 2-point functions of a CFT operator ${\cal O}$ of dimension $\Delta_\pm$, and also that the quantity $\tilde J$ corresponds to the \emph{source} of the boundary CFT operator. The general case in which $\Delta_{\cal O}$ is arbitrary will be analyzed in Section~\ref{double}.

We start with the choice $\Delta_{\cal O}=\Delta_-$. In this case one can see that the undeformed CFT action, i.e. eq~(\ref{reg}) with $\tilde J=0$, is finite as $\epsilon\rightarrow0$ for $0<\nu<1$. In the derivation of the 2-point function we will thus assume that the relation $0<\nu<1$ is satisfied. By Fourier transforming the scalar, the on-shell regularized action becomes
\ba\label{on-shell}
I_{reg}\Big{|}_{on-shell}=\frac{1}{2}\int\frac{d^dq}{(2\pi)^d}\tilde J(\epsilon,-q)\epsilon^{-d}\phi(\epsilon,q).
\ea
Finiteness of the action in the presence of the source term forces the normalization of $\tilde J$. This latter condition translates into 
\ba\label{source}
\tilde J(\epsilon,q)&=&2\nu \epsilon^{\Delta_+}J(q)\\\no
&=&\epsilon^{d}(2\nu\epsilon^{-\Delta_-}) J(q), 
\ea
where the numerical factor is conventionally chosen so that the boundary condition~(\ref{bc}) reads $\beta=J$. The solution of the Euclidean equations of motion satisfying~(\ref{bc}) with our choice $\Delta_{\cal O}=\Delta_-$ and regular in the interior is
\ba\label{phi}
\phi(q,z)=-\tilde J\left(\frac{z}{\epsilon}\right)^{d/2}\frac{K_\nu(qz)}{q\epsilon K_{\nu-1}(q\epsilon)},
\ea
with $K_\nu(x)$ being the modified Bessel function of the second kind. Using the mapping~(\ref{source}), the on-shell action finally becomes
\ba\label{os1}
I_{reg}\Big{|}_{on-shell}=-\frac{1}{2}\int \frac{d^dq}{(2\pi)^d} J(-q)G_-(q) J(q),
\ea
where
\ba\label{2}
G_-(q)=\epsilon^{2\nu}\frac{(2\nu)^2K_\nu(q\epsilon)}{q\epsilon K_{\nu-1}(q\epsilon)}
\ea
Recalling that the condition $\Delta_{\cal O}=\Delta_-$ admits a consistent gravitational description for $\nu<1$, we can now consider the continuum limit $\epsilon\rightarrow0$. From the small argument expansion of the Bessel function we find
\ba
\frac{K_\nu(x)}{x K_{\nu-1}(x)}=\frac{2^{2\nu}}{2\nu}\frac{\Gamma(1+\nu)}{\Gamma(1-\nu)}x^{-2\nu}+\dots
\ea
where the dots are subleading in $x\rightarrow0$ if $\nu<1$. Retaining  only the finite part, the 2-point function~(\ref{2}) reduces to
\ba\label{G-}
G_-(q)\equiv+2\nu\frac{\Gamma(1+\nu)}{\Gamma(1-\nu)}\left(\frac{q}{2}\right)^{-2\nu},
\ea
in agreement with the results of~\cite{Klebanov}. This is the correct 2-point function for a CFT operator with dimension $\Delta_-$. In the range $0<\nu<1$ the correlator satisfies the unitarity bound $Im[G_-]\geq0$, if evaluated on the physical cut~\cite{cw}.

In a similar way we can derive the correlator for a dual operator of dimension $\Delta_+\geq d/2$ by choosing the boundary mass term to be $\Delta_{\cal O}=\Delta_+$. In this case the undeformed CFT action is regular as $\epsilon\rightarrow0$ if $\nu\geq0$, and we will assume that this condition is satisfied. Normalizing the source so that $\alpha=J$ we have
\ba\label{G+}
G_+(q)&=&-2\nu\frac{\Gamma(1-\nu)}{\Gamma(1+\nu)}\left(\frac{q}{2}\right)^{2\nu}.
\ea
In the derivation of the latter expression we only kept the leading nonanalytic term in $q$. The discarded analytic terms correspond to local divergences. As expected from the field theory perspective, such divergences are absent if the CFT operators has dimension $\Delta_-<d/2$, in agreement with our derivation of~(\ref{G-}). The calculation leading to~(\ref{G+}) simply assumed $\nu\geq0$; in that range the unitarity bound $Im[G_+]\geq0$ is satisfied. The degenerate case $\nu=0$ requires a slightly different normalization of the source in order to obtain a finite correlator~\cite{Klebanov}, see also~(\ref{source}).

An important result of the present discussion -- which will be crucial in what follows -- is summarized by the mapping~(\ref{source}) between the boundary coupling $\tilde J$ and the field theoretic source $J$ of the CFT operator dual to $\phi$. In eq.~(\ref{source}) we assumed the operator has dimension $\Delta_-$, but a similar formula holds if $\Delta_{\cal O}=\Delta_+$.

\subsection{Multitrace deformations}
\label{multi}

In the previous subsection we saw that the generating functional for an operator ${\cal O}$ of conformal weight $\Delta_+$ is
\ba\label{Io}
I[\alpha]=-\frac{1}{2}\int \alpha G_+\alpha+\dots,
\ea
where $G_+$ is given above and the dots stand for higher powers of the source $J=\alpha$, which we can neglect in the leading $1/N$ approximation. Similarly, the generating functional of a dimension $\Delta_-$ operator is
\ba
F[\beta]=-\frac{1}{2}\int \beta G_-\beta,
\ea
where now $\beta=J$ denotes the source associated to the dual operator. 

Interestingly, $G_\pm$ are not independent. From~(\ref{G-}) and~(\ref{G+}) we indeed see that
\ba
G_-=- \frac{(2\nu)^2}{G_+},
\ea
where the negative sign comes from the unitarity requirement that the position space correlators are positive definite (in Euclidean space). With this crucial relation at hand, we see that the Legendre transform of the functional $F[\beta]$ with respect to $2\nu\alpha$, namely
\ba
F[\beta]-\int(2\nu\alpha)\beta
\ea
evaluated on the solution of $2\nu\alpha\equiv\frac{\delta F}{\delta\beta}$, gives exactly $I[\alpha]$; in other words, $\beta$ and $2\nu\alpha$ are conjugate variables~\cite{Klebanov}. Recalling the relation between the generating functional and the 1PI action, we see that if the coefficient $\beta$ of~(\ref{sol}) is identified with the CFT source $J$ of the operator ${\cal O}$, $2\nu\alpha$ represents its vacuum expectation value. We will prove this by direct computation in a subsequent section.

The main conclusion of this brief digression is that the on-shell gravity action $I[\alpha]$ must be viewed as the effective 1PI action for the classical field $\langle{\cal O}\rangle=2\nu\alpha$ when the operator ${\cal O}$ has dimension $\Delta_-$. In particular, any deformation of the boundary lagrangian in the regularized theory~(\ref{reg}) must be interpreted as an \emph{explicit} deformations of the effective action for an operator ${\cal O}$ with scaling dimension $\Delta_-$. The same conclusion was previously presented in~\cite{Porrati}. This fact will allow us to describe arbitrary CFT deformations, as we will see shortly.

Now, if we denote with $I[\langle{\cal O}\rangle]$ the effective action for the classical field $\langle{\cal O}\rangle$ -- with the operator having scaling dimension $\Delta_-$ --, a multitrace deformation $\delta{\cal L}=W({\cal O})$ of the bare action (see~(\ref{theory})) is described by the following 1PI functional
\ba\label{If}
I_f[\langle{\cal O}\rangle]=I[\langle{\cal O}\rangle]+\int W(\langle{\cal O}\rangle),
\ea
where $f$ labels the parameters entering $W$. Following our conventions we now denote with $\beta$ the source of the undeformed theory, 
\ba
\beta=-\frac{\delta I}{\delta\langle{\cal O}\rangle}.
\ea
In the presence of the deformation this identification is lost. The source of the deformed theory, $J$, is related to $\beta$ via
\ba
J\equiv-\frac{\delta I_f}{\delta\langle{\cal O}\rangle}=\beta-\frac{\delta W}{\delta\langle{\cal O}\rangle}.
\ea
Equivalently, we can imagine that the deformation $W$ already includes the source term. With this convention the above relation reads
\ba\label{Jf}
\beta=\frac{\delta W}{\delta\langle{\cal O}\rangle}.
\ea
Because $\langle{\cal O}\rangle$ is ultimately connected to $\alpha$, we see that the presence of a multitrace deformation is parametrized by a relation between the two coefficients $\alpha,\beta$ appearing in~(\ref{sol})~\cite{Witten}\cite{SS}.

In complete generality, consider a regularized action of the form
\ba\label{I'}
I_{reg}&=&\frac{1}{2}\int d^dx\int dz\, \sqrt{g}\left[(\nabla\phi)^2+m^2\phi^2+\dots\right]\\\no
&+&\int d^dx \sqrt{h}\left[U(\phi)\right]_{z=\epsilon},
\ea
where the dots stand for arbitrary interactions and higher derivatives, whereas $U$ is an arbitrary functional density of $\phi$ -- typically including a source term $\tilde J\phi$. If we define 
\ba\label{W}
\tilde W(\phi)=U(\phi)-\frac{\Delta_-}{2}\phi^2,
\ea
the boundary condition~(\ref{bc}) generalizes to
 \ba\label{Jf'}
\left[z\phi'-\Delta_-\phi\right]_{\epsilon}=\left[\tilde W'\right]_{\epsilon},
\ea
where from now on a prime denotes derivative with respect to the argument. We claim that~(\ref{Jf'}) represents the gravity dual of~(\ref{Jf}): eq.~(\ref{Jf'}) enforces the CFT constraint~(\ref{Jf}) as a functional relation between the coefficients $\alpha,\beta$ of~(\ref{sol}), and therefore systematically implements the boundary conditions discussed in~\cite{Witten}. As anticipated, the CFT deformation is mapped into the boundary action $U$.

In general, the couplings of $U$, and ultimately of $\tilde W$, depend on the cutoff. We will see this explicitly for the specific case of a quadratic deformation in Section~\ref{double}, but the result is general (see also~\cite{Sundrum}). It will be clear from our analysis that this dependence on the RG scale $\epsilon$ is crucial in obtaining the CFT formula~(\ref{Jf}) from~(\ref{Jf'}). 

We can finally formalize our claim stating that a generic CFT deformation $W$, see~(\ref{theory}), is mapped into the boundary functional~(\ref{W}) via the identification
\ba\label{result}
\tilde W(\phi)\equiv\epsilon^d W(2\nu\alpha),
\ea
where the equivalence is valid for $\epsilon\rightarrow0$.


\section{Doubletrace deformations}
\label{double}

In this section we will address the physical meaning of a generic boundary mass term $\Delta_{\cal O}$ in~(\ref{reg}), and verify the validity of our identification~(\ref{result}) on the explicit, tractable example of doubletrace deformations. The RG evolution of the boundary couplings will be compared to the field theory result, and a precise mapping for the CFT violating parameter $a$ of~(\ref{Beta}) will be given.

\subsection{The RG flow}
\label{RG}

We begin our discussion by emphasizing that our regularization $\epsilon\neq0$ was introduced in Section~\ref{systematic} as a computational artifact, and that no physical meaning is encoded in $\epsilon$. In particular, the gravity theory should not depend on the regulator. Let us see what the implications of this request are. Define  
\ba\label{Deltaf}
\Delta_{\cal O}=\Delta_-+\bar f.
\ea
In terms of the new variable $\bar f$, and using the identification~(\ref{source}), the boundary condition~(\ref{bc}) (or, more generally, eq.~(\ref{Jf'})) becomes
\ba\label{BC}
\beta=J+\frac{\bar f}{2\nu}\left(\alpha\epsilon^{-2\nu}+\beta\right).
\ea

Let us first focus on the dynamics of the CFT in the absence of the external source $J$. In this case~(\ref{BC}) simplifies to
\ba\label{f}
\bar f=\frac{2\nu}{1+\frac{\alpha}{\beta}\epsilon^{-2\nu}}.
\ea
Requiring that the gravity theory does not depend on $\epsilon$ is tantamount to saying that $\alpha,\beta$ do not depend on the UV cutoff. This can only be compatible with the boundary condition if $\bar f$ has a dependence on $\epsilon$ expressed by:
\ba\label{barf}
\epsilon^{-1}\frac{d\bar f}{d\epsilon^{-1}}=\bar f^2-2\nu\bar f.
\ea
This result generalizes the analysis of the $\nu=0$ case performed in~\cite{Witten}.

The beta function~(\ref{barf}) is precisely the RG flow of the coupling of the doubletrace deformation~(\ref{Beta}) when $a=0$, namely when the undeformed theory is an actual CFT. We thus argue that the doubletrace deformation $W=f{\cal O}^2$ is mapped into a boundary term $\bar f\phi^2$ in the gravity theory, and that the coupling $\bar f$ is in one to one correspondence with the renormalized and dimensionless coupling of the field theory. This agrees with our general recipe~(\ref{result}).

The explicit relation between $f$ and $\bar f$ will be derived below. For now we observe that the critical exponents derived by~(\ref{barf}), namely
\ba
\beta'_{\bar f}=\pm2\nu,
\ea
coincide with the ones found on the field theory side, see~(\ref{Beta}) with $a=0$, if we identify $\nu=\sqrt{D}$.

We therefore see that the choices $\Delta_{\cal O}=\Delta_-,\Delta_+$ (that is $\bar f=0,2\nu$) discussed in Section~\ref{two} correspond respectively to an UV and an IR fixed point for the deformed CFT (that is for the coupling $\bar f$). At the fixed points we have either $\beta=0$ or $\alpha=0$ depending on the fixed point being UV or IR.

\subsection{The 2-point function}

The generating functional in the presence of the deformation $\bar f\neq0$ (arbitrary $\Delta_{\cal O}$) can be derived along the lines presented in Section~\ref{systematic}, the only difference consists in a modification of the boundary conditions. Imposing the constraint~(\ref{bc}) -- where $\Delta_{\cal O}$ is now the generic expression~(\ref{Deltaf}) -- on the solution of~(\ref{EOM}) we generalize~(\ref{phi}) to the $\bar f\neq0$ case:
\ba
\phi(q,z)=-\tilde J\left(\frac{z}{\epsilon}\right)^{d/2}\frac{K_\nu(qz)}{q\epsilon K_{\nu-1}(q\epsilon)+\bar f K_{\nu}(q\epsilon)}.
\ea
Thanks to the identification~(\ref{source}), the regularized on-shell action~(\ref{os1}) now generalizes to
\ba
I_{reg}\Big{|}_{on-shell}=-\frac{1}{2}\int \frac{d^dq}{(2\pi)^d} J(-q)G_{\bar f}(q) J(q),
\ea
where we defined
\ba\label{Gf}
G_{\bar f}(q)=\frac{G_-(q)}{1+f G_-(q)},
\ea
and
\ba\label{ident}
\bar f&=&(2\nu)^2\epsilon^{2\nu} f\\\no
&=&\epsilon^d\left(2\nu\epsilon^{-\Delta_-}\right)^2 f.
\ea
The numerical factor $(2\nu)^2$ in~(\ref{ident}) reflects our conventions in the normalization of the Green's function, and hence of the source $\tilde J$~(\ref{source}).

The expression~(\ref{Gf}) may have been obtained from a Dyson series of the form $G_{\bar f}=G_-+(-f)G_-(-f)+\dots$, corresponding to the correlator of a CFT deformed by the doubletrace operator $W=f{\cal O}^2/2$~\cite{Witten}\cite{GK}. This allowed us to make the identification~(\ref{ident}) between the boundary coefficient $\bar f$ and the (bare) coupling $f$. 

The correlator~(\ref{Gf}) describes a flow between two fixed points of the renormalization group, as~(\ref{barf}) anticipates. In fact, the leading nonanalytic expression for low or high momenta is
\ba
G_{\bar f}\sim&\left\{ \begin{array}{ccc}  -q^{+2\nu}\quad\quad q\ll f^{1/(2\nu)}\\
+q^{-2\nu}\quad\quad q\gg f^{1/(2\nu)}. \end{array}\right.
\ea
From the scaling behavior of the Green's function we confirm that $\Delta_-$ and $\Delta_+$ are the UV and IR dimensions of ${\cal O}$, respectively.

We have just shown that the $AdS_{d+1}$ background is capable of reproducing the 2-point function of the deformed ($\bar f\neq0$) field theory. This suggests that pure $AdS_{d+1}$ is the correct geometry on which to compute correlation functions in the presence of doubletrace deformations. This conclusion seems counterintuitive, because we are used to think that a deformation of the CFT should be mapped into a deformation of the dual geometry. Yet, we will prove the nonperturbative classical stability of $AdS_{d+1}$ in the presence of the doubletrace deformation in Section~\ref{stability}. Only when subleading corrections (quantum effects on the gravity theory) are taken into account the RG flow is found to actually induce a modification of the geometry~\cite{GM}.

We conclude this section with an explicit test of our formula~(\ref{result}). In the case of $W=J\langle{\cal O}\rangle+f\langle{\cal O}\rangle^2/2$ the field theory arguments of Section~\ref{multi} imply $\beta=J+f\langle{\cal O}\rangle$, see eq.~(\ref{Jf}). If the regularized theory~(\ref{reg}) aspires to describe this physics, it should be able to reproduce this result. To show that this is indeed the case we first have to relate the vacuum expectation value $\langle{\cal O}\rangle$ of the undeformed theory to the coefficient $\alpha$ of~(\ref{sol}). Inserting the asymptotic form~(\ref{sol}) in the on-shell action, imposing the boundary constraint~(\ref{BC}), and deriving the action with respect to $J$ ($\bar f$ does not depend on $J$), gives
\ba\label{vac}
\langle{\cal O}\rangle_J=2\nu\left(\alpha+J\frac{d\alpha}{dJ}\right),
\ea
for $\epsilon\rightarrow0$. We thus recover the well known statement that the vacuum of the theory in the absence of the deformation ($\bar f=J=0$) is given by $\langle{\cal O}\rangle=2\nu\alpha$. Now, in the limit $\epsilon\rightarrow0$ and using~(\ref{ident}) one sees that our boundary condition~(\ref{BC}) reads $\beta=J+f(2\nu\alpha)$, which is indeed the expected result~(\ref{Jf}).

More explicitly, with the help of~(\ref{source}) and~(\ref{ident}) we see that, in the limit $\epsilon\rightarrow0$,
\ba\label{test}
\tilde W=\tilde J\phi+\frac{\bar f}{2}\phi^2\rightarrow \epsilon^d\left[J(2\nu\alpha)+\frac{f}{2}(2\nu\alpha)^2\right].
\ea
This result confirms our map~(\ref{result}). The extra factor $\epsilon^d$ in~(\ref{result}) exactly cancels the $\sqrt{h}$ in~(\ref{I'}), and ensures that the deformation survives as the cutoff is removed. Notice that the running of the couplings $\tilde J,\bar f$ is crucial in deriving~(\ref{test}).

\subsection{The role of nonminimal couplings to gravity}
\label{nonmin}

In a generic theory there is no symmetry forbidding nonminimal couplings to gravity. In particular, we expect a term of the form $\xi\phi^2 R/2$ to arise in a non-supersymmetric setup. Because this term is quadratic in the scalar, we anticipate some modifications of the 2-point function already at leading order in the planar limit. Let us consider the following action (now in Minkowski space):
\ba\label{HG}
I_{reg}&=&\frac{1}{2}\int d^{d+1}x\sqrt{g}\left((\nabla\phi)^2-m^2\phi^2+\xi R\phi^2\right)\\\no
&+&\frac{1}{2}\int d^dx\sqrt{h}\left[2\xi K\phi^2-(\bar f+\bar\Delta_-)\phi^2\right]_{z=\epsilon}.
\ea 
We have dropped the source term because we would like to focus on the dynamics~(\ref{theory}) with $W$ a doubletrace deformation. 

The boundary term $\sqrt{h}\xi K\phi^2$ is a generalization of the Hawking-Gibbons term, and it is essential in order to formulate a consistent variational problem in the presence of a nonminimal coupling (see the Appendix~\ref{A}). Here, $K=\nabla_a n^a$ is the trace of the extrinsic curvature and $n^a$ is an unit vector normal to the boundary. In our case $n^a=(0,0,\dots,0,z)$ is space-like and we find
\ba
K=-d.
\ea
The additional, $\propto\xi$, boundary term has a crucial impact on the resulting physics. This is readily understood by implementing the standard AdS/CFT prescription. The effect of nonminimal couplings to gravity were also considered in~\cite{Minces2}.

Here, we find it convenient to analyze the implications of a nonminimal coupling by means of our modified prescription~(\ref{Jf'}). For this purpose we have included in~(\ref{HG}) the boundary $\bar\Delta_-\phi^2/2$ operator, where 
\ba\label{below}
\bar\Delta_-&=&\frac{d}{2}-\bar\nu\\\no
\bar\nu&=&\sqrt{\bar m^2+\frac{d^2}{4}}
\ea 
includes an effective bulk mass term $\bar m^2=m^2-\xi R$. Yet, this choice, i.e. the definition of $\bar f$ in~(\ref{HG}), is completely arbitrary, and the discussion that follows does not depend on it. With our convention, the boundary condition~(\ref{Jf'}), for $\tilde J=0$, becomes:
\ba
z\phi'-\bar\Delta_-\phi=(\bar f+2d\xi)\phi
\ea
and can be equivalently expressed as
\ba\label{fxi}
\bar f+2d\xi=\frac{2\bar\nu}{1+\frac{\alpha}{\beta}\epsilon^{-2\bar\nu}}.
\ea 
This generalizes eq.(\ref{f}) to the $\xi\neq0$ case. If we insist that our theory does not depend on the regularization, we are again forced to conclude that there exists a scale dependence in the deformation $\bar f(\epsilon)$~\footnote{The nonminimal coupling $\xi$ clearly does not depend on the cutoff $\epsilon$, and so do the mass $m^2$ and any bulk parameter.}: 
\ba\label{xi}
\epsilon^{-1}\frac{d\bar f}{d\epsilon^{-1}}=\bar f^2+2(2d\xi-\bar\nu)\bar f+(2d\xi-2\bar\nu)2d\xi.
\ea
This is the dual of~(\ref{Beta}) with $a\neq0$. Computing the discriminant of~(\ref{xi}) we find
\ba\label{D'}
D=(2d\xi-\bar\nu)^2-(2d\xi-2\bar\nu)2d\xi=\bar\nu^2,
\ea
which is the identification anticipated in the introduction.

In order to appreciate the physical content of the present conclusion it is worth focussing on an explicit example. Consider type IIB supergravity. It is known that a consistent truncation of this theory at the 2-derivative level does not contain a nonminimal coupling, i.e. $\xi=0$. However, as soon as supersymmetry is violated, there is no reason to expect the previous condition to hold, and generally $\xi\neq0$. What we have shown above is that the introduction of a nonminimal coupling necessarily requires a deformation of the boundary physics. This deformation is parametrized by a doubletrace deformation of the dual CFT with coupling $\bar f\neq0$ and beta function~(\ref{xi}). 

This picture has a straightforward dual interpretation. As long as supersymmetry is preserved, type IIB supergravity is dual to ${\cal N}=4$ SYM at large $N$ and large 't Hooft coupling, which is known to be exactly conformal. If supersymmetry is broken, the would-be ${\cal N}=4$ SYM must be supplemented with doubletrace deformations ($\bar f\neq0$) in order to obtain a consistent quantum field theory. The beta function for the coupling of the deformation acquires the form~(\ref{Beta}) -- where $\Delta$ is the dimension of ${\cal O}$ when $\bar f=0$ and corresponds to our $\Delta_-=d/2-\nu$, while $a$ measures the deviation from the supersymmetric scenario. 

We therefore conclude that the parameter $a$ is mapped into the quantity $\xi$ by the gauge/gravity correspondence. Matching the discriminant~(\ref{D'}) with the field theoretic~(\ref{D}) we have
\ba\label{a}
\left(\Delta-\frac{d}{2}\right)^2-va=\frac{d^2}{4}+m^2-\xi R,
\ea
which allows us to formally relate the nonminimal coupling $\xi$ to the CFT violating term $a$ as $\xi R=va$.

\section{The gravity spectrum}
\label{RES}

We would like to analyze the spectrum of the scalar fluctuations of~(\ref{reg}); the generalization to the nonminimally coupled case $\xi\neq0$, eq.~(\ref{HG}), is straightforward. This study will enable us to identify a condition for the perturbative stability (absence of tachyons) of the $AdS_{d+1}$ background in the presence of doubletrace deformations. A nonperturbative (though classical) condition for stability in the presence of arbitrary deformations $W$ and nonminimal couplings will be presented in Section~\ref{stability}.

In agreement with the field theory expectations reviewed in the introduction, we will see that a resonant mode develops in the conformal window $0<\nu<1$. This resonance will be argued to be a generic feature of models interpolating between two fixed points of the RG flow.

\subsection{Physical excitations}

By defining $\phi=z^{\frac{d-1}{2}}\psi$, the equation of motion~(\ref{EOM}) reduces to a Schroedinger form
\ba\label{Seq}
-\psi_p''+\frac{\nu^2-\frac{1}{4}}{z^2}\psi_p=p^2\psi_p,
\ea
where $p^2=-q^2$ is the momentum in Minkowski space. Equation~(\ref{Seq}) is equivalent to a quantum mechanical problem for a particle in a central potential. 

When $\nu$ is complex, i.e. $m^2<m_{BF}^2$, equation~(\ref{Seq}) admits tachyonic solutions and the hamiltonian becomes unstable. This does not represent a problem in a nonrelativistic setup, in which $p^2=E<0$ is allowed~\cite{Moroz}, but it certainly does in our case. The analogy with the quantum mechanical system suggests that, in order to obtain a positive vacuum energy when the scalar violates the Breitenlohner-Freedman bound~\cite{BF}, the system has to be enclosed in a box. This translates into the introduction of an IR and an UV boundaries and signals a breaking of conformality~\cite{CL'}\cite{CL}. From now on we will focus on the case $0<\nu<1$, i.e. on the conformal window of the dual CFT.

The wavefunctions $\psi_p$ must satisfy appropriate boundary conditions. The boundary conditions at the conformal boundary, see~(\ref{bc}), are imposed on general field configurations, in particular on the physical excitations. In terms of the field $\psi_p$ these are~\footnote{We are focussing on the theory~(\ref{theory}) with $W=f{\cal O}^2$, and hence we assume that $\tilde J=0$.}
\ba\label{blah}
z\psi_p'+\left(\nu-\bar f-\frac{1}{2}\right)\psi_p=0.
\ea
The boundary conditions at $z\rightarrow\infty$ follow from normalizability (or equivalently from the second equation in~(\ref{bcon})) and, in terms of the rescaled field $\psi_p$, reduce to the requirement of square integrability in the region $z>0$. Because the Schroedinger potential in~(\ref{Seq}) approaches zero in the far IR, we deduce that the spectrum contains a continuum of delta-function normalized profiles $\psi_p$ of masses $p^2>0$.

Tachyonic solutions are also present for generic $\bar f<0$. The normalizable solution for $p^2<0$ is given by $\psi_T=N_T\sqrt{z}K_\nu(p_Tz)$, where $N_T$ is determined by the normalization, while the eigenvalue is given by the condition~(\ref{blah}) at the UV cutoff $\epsilon$:
\ba
p_T\epsilon\frac{K_{\nu-1}(p_T\epsilon)}{K_{\nu}(p_T\epsilon)}=-\bar f.
\ea
Using~(\ref{2})(\ref{Gf})(\ref{ident}) the above relation is seen to coincide with the condition for the existence of an Euclidean pole in the 2-point function~(\ref{Gf}), as one would have expected. The $AdS_{d+1}$ space is therefore unstable if $\bar f<0$, and consistently we see that in this case the field theoretic beta function~(\ref{Beta}) develops an IR Landau pole. As $\bar f\rightarrow0$ the tachyon becomes lighter and at $\bar f=0$ it interpolates a massless bound state with profile
\ba
\psi_0=N_0 z^{1/2-\nu}.
\ea
This state is not normalizable ($N_0=0$) in the conformal window $0<\nu<1$, and thus has no physical impact. Indeed, no such pole is found in~(\ref{Gf}).

For $\bar f\geq0$ there are no tachyons in the spectrum, and the theory is now perturbatively stable. We show below that in this case the tachyon is converted into a resonant mode.

\subsection{Resonant pole}

The analogy between the Schroedinger equation~(\ref{Seq}) and the nonrelativistic model for a particle scattering off a central potential suggests to look for resonant modes (i.e. Gamow states) among solutions of the eigenvalue problem that satisfy pure outgoing boundary conditions at radial infinity, namely $\psi\sim e^{ipz}$ at $z\gg1$. This modified IR boundary condition selects the solution 
\ba\label{res}
\psi_R=\sqrt{z}H^{(1)}_\nu(p_Rz),
\ea
where $H^{(1)}_\nu=J_\nu+iY_\nu$ is the Hankel function.
By imposing the boundary condition~(\ref{blah}) at the conformal boundary we find
\ba\label{R}
p_R\epsilon\frac{H^{(1)}_{\nu-1}(p_R\epsilon)}{H^{(1)}_{\nu}(p_R\epsilon)}=\bar f.
\ea
Equation~(\ref{R}) admits solutions $p_R^2$ for $\bar f\geq0$ only. Taking the limit $\epsilon\rightarrow0$ and using the identification~(\ref{ident}), the eigenvalue equation can be rewritten as:
\ba\label{pole}
(-p_R^2)^\nu&=&-\epsilon^{-2\nu}\frac{2^{2\nu}}{2\nu}\frac{\Gamma(1+\nu)}{\Gamma(1-\nu)}\bar f\\\no
&=&-2\nu  \frac{\Gamma(1+\nu)}{\Gamma(1-\nu)} 2^{2\nu} f.
\ea
This is precisely the condition for the appearance of a pole $p_R^2$ in~(\ref{Gf}). In the case $\bar f>0$, as opposed to $\bar f<0$, this pole is generally complex.

Explicit solutions of the resonant condition have been derived in~\cite{cw}. In that paper it was shown that a single pole out of the infinite solutions of~(\ref{pole}) has the correct status of resonance, and that this resonance may be seen as the remnant of the Nambu-Goldstone mode of conformal invariance found in the nonsymmetric phase. Notice that in passing from the first to the second line of~(\ref{pole}) it is crucial that $\bar f$ is a relevant coupling. If $\bar f$ is irrelevant, or even marginal, then the pole $p_R^2$ would diverge as some negative power of $\epsilon$ and the resonance would decouple. We conclude that a resonance emerges only if the coupling $\bar f$ is defined in the range $0<\bar f<2\nu$, in which the theory flows between two CFT's, in perfect agreement with the field theory analysis~\cite{cw}.

Resonant states are found by giving up the hermitianity of the Schroedinger operator~(\ref{Seq}). As a result, their mass is generally complex, and their profiles are not normalizable. Specifically, due to its non-vanishing width the resonant profile~(\ref{res}) asymptotically reads:
\ba
\psi_R\sim e^{ip_Rz}=e^{iMz+\frac{\Gamma}{2}z},
\ea
where we introduced the definition $p_R=M-i\Gamma/2$. The resonance is thus generally peaked in the IR if $\Gamma>0$, as its compositeness nature anticipates, and becomes more and more overlapped with the UV region as the width vanishes. It is easy to see when this limiting case occurs. From~(\ref{pole}) we find that the ratio between the width $\Gamma$ and the mass $M$ of the resonance is:
\ba
\frac{\Gamma}{M}=-2\tan\left(\frac{1+2k+\nu}{2\nu}\pi\right),
\ea
where $k=0,\pm1,\pm2,\dots$. A small width is attained for $\nu\rightarrow1^-$ and if $k=-1$~\cite{cw}. In the latter case we have
\ba
\frac{\Gamma}{M}\simeq\frac{1-\nu}{\nu}\pi,
\ea
indicating that the resonance width vanishes as the dual operator ${\cal O}$ interpolates a free scalar field.

We emphasize that such an IR-localized mode does not back-react on the geometry. Indeed, this state does not satisfy the boundary conditions~(\ref{bcon}) imposed in the formulation of the variational problem and therefore does not contribute to the equations of motion. Because of its off-shell nature, the resonance~(\ref{res}) has no impact on the vacuum structure, even though its profile is formally divergent in the IR. We will have more to say on the stability of the $AdS_{d+1}$ geometry in the next section.

The emergence of a resonant mode in the presence of a boundary mass term is a generic feature. This can be illustrated by studying the flat space-time limit. Consider the equation of motion for $\phi$ in the limit in which the curvature vanishes:
\ba
-\phi''+m^2\phi=p^2\phi,
\ea
where $p^2$ is the Minkowski momentum and $m^2$ the bulk mass. Suppose now that the space-time is confined in the region $z>0$, or similarly that $-\infty<z<\infty$ with a $Z_2$ orbifold symmetry around the origin. Let us place a brane at $z=0$ with a localized mass term of the form $\frac{\Delta}{2}\phi^2$. The equation of motion for $\phi$ now admits an exponentially localized solution with momentum $p^2=m^2-\Delta^2$. Absence of tachyons then requires $m^2\geq\Delta^2$, which is basically the stability condition $\bar f\geq0$ of our model~(\ref{Seq}). Once gravity is switched on, one expects the localized mode to decouple because of the curvature-induced suppression of the potential~(\ref{Seq}) at large $z$. Yet, by continuity of the physical observable, a remnant of the long range force mediated by the localized mode should be found. This is precisely the resonant mode~(\ref{res}).

The previous argument can be generalized to more involved frameworks. We would like to convince the reader that a resonant behavior of the type discussed above is characteristic of geometries interpolating between two $AdS$ spaces. Our argument is very simple, and rather general. A deformation of the gravitational background should be induced by the condensation of a set of fields which we collectively call $\phi$. The assumption that the geometry interpolates between two $AdS$ spaces is equivalent to the statement that $\phi$ approaches constant values for $z\rightarrow0,\infty$. If we now assume that gravity can be switched off -- for example by sending the Planck mass to infinity -- in such a way that the previous asymptotic condition on the scalar is not drastically changed, i.e. in such a way that $\phi$ still interpolates between constant values (which may differ from the gravity model), we infer that the spectrum contains a normalizable massless scalar mode. In fact, in the absence of gravity our assumptions on the scalar vacuum imply that translational invariance has been spontaneously broken, and that an associated normalizable Nambu-Goldstone mode must appear. In analogy to the exponentially localized mode discussed in the above example, its presence is expected to be signaled by a resonant behavior in the gravitational model.

A class of models with this property has been studied in~\cite{mio}, and consists in those theories in which the scalar potential $V$ can be written in terms of a function $P(\phi)$ as $2V=P'^2-\frac{dG}{d-1} P^2$, with $G$ the Newton constant (see also Section~\ref{stability}). Background configurations for $\phi$ satisfying $d$-dimensional Poincare invariance can be written as solutions of the first order equation $\phi'=P'(\phi)$, and hence persist unmodified in the limit $\mu^{d-1}G\rightarrow0$ in which gravity decouples. As gravity is continuously switched on, the Nambu-Goldstone mode of translational invariance is turned into a scalar resonance manifesting itself in the 2-point function of the bulk scalar $\phi$~\cite{mio}. 

Recalling the gauge/gravity relation between the coordinate $z$ and the RG scale $\mu$, we associate the resonance found in these models with a would-be Nambu-Goldstone mode of dilatation invariance, in agreement with the field theoretic arguments of~\cite{cw}.

\section{Stability of AdS/CFT}
\label{stability}

The analysis of the spectrum of~(\ref{reg}) has shown that perturbative stability of the $AdS_{d+1}$ geometry in the presence of doubletrace deformations is ensured as far as $\bar f\geq0$, in agreement with the RG analysis. In this section, we sill study the gravitational stability for arbitrary potentials $W$ by means of the nonperturbative approach of~\cite{des2}\cite{des3}\cite{des4}.

We will consider scalar/gravity systems supporting both regular and irregular terms, namely theories in which both the $\alpha,\beta$ coefficients of~(\ref{sol})(\ref{sol'}) are in principle present. The stability of systems satisfying $\alpha=0$ have been studied at length in the literature starting with~\cite{Boucher}. The stability for minimally coupled scalar/gravity systems with $\alpha\neq0$ has been discussed in~\cite{des2}\cite{des3}\cite{des4}. Our main aim is to generalize the results of~\cite{des2}\cite{des3}\cite{des4} to models with nonminimal couplings to gravity. In addition, we will provide a gauge/gravity interpretation for the requirement imposed in~\cite{des2}\cite{des3}\cite{des4} that the bulk scalar potential is determined by a ``superpotential".

\subsection{Stability of the minimally coupled system}

Consider the generalization of~(\ref{reg})
\ba\label{grav}
I&=&\int_M d^{d+1}x\sqrt{ g}\left[-\frac{1}{2G} R+\frac{1}{2}( \nabla\phi)^2-V\right]\\\no
&+&\int_{\partial M} d^dx\sqrt{ h}\left[- \frac{1}{G}K- U\right],
\ea
where $U$ and $V$ are functions of $\phi$ and $G$ is the Newton constant. A correct formulation of this theory requires the definition of appropriate boundary conditions. Following the prescription developed in Section~\ref{multi} we choose
\ba\label{bc''}
n^a\nabla_a\phi=U'\quad\quad \delta g_{ab}=0.
\ea
The equations of motion read
\ba\label{EOM'}
-\nabla^2\phi&=&V'\\\no
\frac{1}{G}\left(R_{ab}-\frac{1}{2}g_{ab}R\right)&=&\nabla_a\phi\nabla_b\phi-g_{ab}\left(\frac{1}{2}(\nabla\phi)^2-V\right).
\ea
In the above equations and below a prime denotes derivative with respect to $\phi$.

In the following discussion we will concentrate on systems that asymptotically approach the $AdS_{d+1}$ space
\ba\label{as}
\phi\rightarrow0\quad\quad ds^2\rightarrow\frac{1}{z^2}\eta_{ab}dx^adx^b.
\ea
Self-consistency then requires that the asymptotic $AdS$ background~(\ref{as}) is a solution of the equations of motion, i.e. $V'(0)=0=U'(0)$ and
\ba\label{L}
G\,V(0)=-\frac{d(d-1)}{2}=-\frac{1}{2}\frac{d-1}{d+1}R.
\ea
We normalized the curvature length to 1 to conform with the convention used throughout the paper.

Given our boundary conditions, we then ask how $\phi=0$ is approached by a generic solution of the equations of motion. In the following analysis we focus on models in which the scalar $\phi$ has a mass in the range $m_{BF}^2<m^2<m_{BF}^2+1$; in this case we expect both fast and slow falloff terms to be generally present. Let us be a bit more precise and assume that~(\ref{grav}) has a $Z_2$ $\phi\rightarrow-\phi$ symmetry. Under this hypothesis the potentials $V,U$ do not contain a $\phi^3$ term and, given the asymptotic form of the metric~(\ref{as}), the leading $z\sim 0$ expression for a scalar $\phi$ for $d\geq4$ is entirely dictated by the quadratic potential 
\ba\label{asphi}
\phi\sim \alpha z^{\Delta_-}+\beta z^{\Delta_+}
\ea
where $\Delta_\pm$ are defined in~(\ref{Dim}) with $m^2=V''(0)$. 

The constraints $0<\nu<1$ and $d\geq4$ imply that $4\Delta_->d$, and ensure that near the conformal boundary $\left(1+O(\phi^2)\right)\phi\approx\phi$. This is basically a statement regarding the irrelevance of the higher order correlation functions of the undeformed CFT. The dual interpretation is that higher order correlators are expected to be suppressed by powers of $1/N$, within the CFT. This does not mean that higher order terms in $U$ are suppressed, though: the couplings entering the CFT deformation $W$ defined in~(\ref{theory}) are in principle arbitrary (i.e. the deformation~(\ref{result}) generally has a nontrivial dependence on the UV cutoff such that higher powers of the field may survive after the continuum limit is taken). In other words, at the UV boundary we expect that
\ba\label{oooo}
V\rightarrow V(0)+\frac{V''(0)}{2}\phi^2,
\ea
with $U$ generic.

As shown in Section~\ref{multi}, the boundary condition induced by an arbitrary $U$ implies a generic relation between $\alpha$ and $\beta$. This typically leads to a breaking of the asymptotic $AdS$ isometries~\footnote{Asymptotic scale invariance is satisfied if the CFT deformation is barely marginal. In our formalism this requires that $\tilde W=\bar\lambda\phi^n$ with $n\Delta_-=d$. In this case~(\ref{Jf'}) and~(\ref{result}) give $\beta=n\lambda(2\nu\alpha)^{\Delta_+/\Delta_-}$, which is precisely the condition ensuring the preservation of the asymptotic $AdS$ isometries found in~\cite{Maeda}.}. Yet, an asymptotic time-like killing vector exists~\cite{Maeda}. Since by diffeomorphism invariance conserved charges associated to space-time symmetries reduce to surface integrals, one concludes that these systems posses a conserved energy which can be used to formulate a stability theorem. Asymptotically $AdS$ spaces with a scalar in the mass range $0<\nu<1$ and with an arbitrary relation $\beta(\alpha)$ have been extensively studied in the context of designer gravity~\cite{des1}.

The relation $\beta(\alpha)$ can be systematically implemented by means of the effective potential~(\ref{W}). In terms of $\tilde W$, and taking into account the asymptotic form of the metric, the boundary condition~(\ref{bc''}) on the scalar becomes~(\ref{Jf'}). Notice that any additive constant in $\tilde W$ is immaterial if the metric satisfies Dirichlet conditions. Without loss of generality we assume that $\tilde W(0)=0$. 

The theory dual to the classical system~(\ref{grav}) with $\tilde W=0$ is a $d$-dimensional CFT at leading order in $1/N$. We hence expect the stability of~(\ref{grav}) to be controlled by the CFT deformation: a sufficient condition for stability of $AdS_{d+1}$ is that $\tilde W\geq0$. A series of papers analyzed the stability of the scalar/gravity system~(\ref{grav}) using a genuinely gravitational perspective~\cite{des2}\cite{des3}\cite{des4}; accordingly, their conclusion is that the energy of the scalar/gravity system is bounded below by the minimum of $\tilde W$, if it exists. See also~\cite{Porrati}. 

In the specific case of a doubletrace deformation we write $\tilde W=\bar f\phi^2/2$, and stability of $AdS$ requires $\bar f\geq0$. This conclusion agrees with the perturbative criteria. Furthermore, it clarifies why the RG analysis of doubletrace deformations presented in Section~\ref{double} that assumed an $AdS_{d+1}$ background was correct at leading order.

The stability theorem established in~\cite{des2}\cite{des3}\cite{des4} applies to non-generic scalar potentials. Specifically, in the proof it is crucial that $V$ be given in terms of a function $P$ as
\ba\label{pot}
2V=P'^2-\frac{dG}{d-1} P^2.
\ea
This is not sufficient, though. One can show that the solutions $P(\phi)$ of~(\ref{pot}) occur in two branches, called $P_\pm$, which differ in particular in the quadratic term in $\phi$ (see~\cite{Papa2} for a systematic study). Near the $AdS$ boundary this means that
\ba\label{Ppm}
P_\pm(\phi)=P(0)+\frac{\Delta_\pm}{2}\phi^2+\dots.
\ea
The stability theorem requires that~(\ref{pot}) admits a global solution $P_-$, the existence of $P_+$ is not sufficient~\cite{des4}. Assuming that a solution $P_-$ of~(\ref{pot}) exists, the theorem says that AdS/CFT is stable provided the minimum of $W$ (or equivalently $\tilde W$) satisfies $W\geq0$.

The necessity of a solution $P_-(\phi)$ can be anticipated by the field theory interpretation. Indeed, recall that the stationary points of the effective action of our $d$-dimensional dual theory satisfy $\beta=W'$, and note that the criteria for the nonperturbative stability of the dual CFT is based on the study of the stationary points of the effective potential $W$. In the gravity language we say that the $AdS_{d+1}$ geometry is stable provided the configurations with $\beta=0$ -- those behaving asymptotically as $\phi=\alpha z^{\Delta_-}$ -- have higher energies compared to pure $AdS_{d+1}$ with $\alpha=\beta=0$. In complete generality one can show that any background manifesting $d$-dimensional Lorentz invariance (the dual boundary theory has no gravity) and having a scalar configuration with the asymptotic form $\phi\propto z^{\Delta_-}$ can be written as a solution of $\phi'=P_-'=\Delta_-\phi+\dots$, with some $P_-$ defined by~(\ref{pot}), see for example~\cite{DeW}. We thus see that the function $P_-$ must exist if our dual CFT interpretation holds.

\subsection{Stability of the nonminimally coupled system}

We would like now to generalize the stability theorem of~\cite{des2}\cite{des3}\cite{des4} to the nonminimally coupled case, the relevance of which has been emphasized in Section~\ref{nonmin}. The field theory suggests that the stability of such systems cannot be simply a consequence of the boundary potential $W$, as it was for the minimally coupled system, because now the undeformed dynamics is not described by a sensible quantum theory. The stability will crucially depend on the interplay between the undeformed theory \emph{and} the boundary deformation.

Limiting ourselves to a 2-derivative action we generalize~(\ref{HG}) to:
\ba\label{NM}
I_{reg}&=&\int_M d^{d+1}x\sqrt{g}\left[-\frac{1}{2G}YR+\frac{1}{2}(\nabla\phi)^2-V\right]\\\no
&+&\int_{\partial M} d^dx\sqrt{h}\left[- \frac{1}{G}YK-U\right],
\ea
The functions $V,U,Y$ depend on $\phi$ in a $Z_2$ invariant way. The meaning of the $YK$ term is elucidated in the Appendix~\ref{A}. The bulk equations of motion for the scalar is the same as in the minimally coupled system, whereas the Einstein equations become
\ba\label{EOMM}
&&\frac{1}{G}\left(R_{ab}-\frac{1}{2}g_{ab}R+g_{ab}\nabla^2-\nabla_a\nabla_b\right)Y\\\no
&&=\nabla_a\phi\nabla_b\phi-g_{ab}\left(\frac{1}{2}(\nabla\phi)^2-V\right).
\ea

The boundary conditions now read 
\ba\label{bc'}
n^a\nabla_a\phi=U'+\nabla_a n^a\frac{Y'}{G}\quad\quad \delta g_{ab}=0.
\ea
The asymptotic forms of $\phi,g_{ab}$ discussed for the minimally coupled system, eq.~(\ref{as}) also apply to the nonminimally coupled case, up to a minimal difference. In fact, the curvature length of the boundary $AdS$ space is still given by~(\ref{L}); however, in the general case $Y''(0)\neq0$ the asymptotic scalar profile~(\ref{asphi}) becomes
\ba\label{ops}
\phi\sim\alpha z^{\bar\Delta_-}+\beta z^{\bar\Delta_+},
\ea
where the exponents $\bar\Delta_\pm$ may contain an effective mass contribution from $Y$. The functional relation $\beta(\alpha)$ is determined by the boundary condition~(\ref{bc'}), in analogy with the minimally coupled case. In~(\ref{ops}) we assumed $0<\bar\nu<1$ and $d\geq4$ and used the fact that, for the same reasons leading to~(\ref{oooo}), near the conformal boundary we can approximate:
\ba\label{bY}
V&\rightarrow& V(0)+\frac{V''(0)}{2}\phi^2\\\no
Y&\rightarrow& 1+\frac{Y''(0)}{2}\phi^2.
\ea
Again, $U$ can be generic.

In order to generalize the stability proof presented in~\cite{des2}\cite{des3}\cite{des4} to the nonminimally coupled model~(\ref{NM}), we find it convenient to perform a change of variables and obtain a minimally coupled system, for which the stability theorem has been derived. The new field variables $\bar\phi(\phi)$ and $\bar g_{ab}(\phi,g)$ are defined by 
\ba\label{cov}
\left(\frac{d\bar\phi}{d\phi}\right)^2=\frac{1}{Y}+\frac{d/G}{d-1}\frac{Y'^2}{Y^2}\quad\quad\bar g_{ab}=Y^{\frac{2}{d-1}}g_{ab}.
\ea
In terms of the new fields, the original action~(\ref{NM}) becomes
\ba\label{new}
I_{reg}&=&\int_M d^{d+1}x\sqrt{\bar g}\left[-\frac{1}{2G}\bar R+\frac{1}{2}(\bar \nabla\bar\phi)^2-\bar V\right]\\\no
&+&\int_{\partial M} d^dx\sqrt{\bar h}\left[-\frac{1}{G}\bar K-\bar U\right],
\ea
where 
\ba
\bar V=\frac{V(\phi)}{Y^{\frac{d+1}{d-1}}},\quad\quad \bar U=\frac{U(\phi)}{Y^{\frac{d}{d-1}}}.
\ea
One can easily check by making use of the equations of motion that the mass $\bar m^2=\bar V''(0)$ of the rescaled field $\bar\phi$ coincides with the effective ($Y$-dependent) mass of the original variable $\phi$ (see below eq.~(\ref{below})):
\ba
\bar V''(0)&=&V''(0)-\frac{d+1}{d-1}V(0)Y''(0)\\\no
&=&V''(0)+\frac{1}{2G}RY''(0).
\ea

Since $Y$ plays the role of an effective Planck mass, we will always assume that $Y>0$. Under this condition the change of variables~(\ref{cov}) is invertible. Furthermore, since the jacobian of the transformation is the unit matrix at the origin, the on-shell physics described by the new system is equivalent to the original, nonminimally coupled theory. It then follows that a stability condition for the minimally coupled system $\bar\phi,\bar g_{ab}$ determines the stability of the nonminimally coupled one.

The nonminimally coupled variables may be referred to as the physical degrees of freedom, to emphasize that the change of variables~(\ref{cov}) is just a mathematical artifact, and it does not have any physical interpretation. The physical fields are believed to be dual to CFT operators, and it is the boundary conditions on these fields, i.e.~(\ref{bc'}), that should be related to the CFT deformation as discussed in Section~\ref{multi}. The minimally coupled system inherits the boundary conditions from the nonminimally coupled variables.

For us the crucial observation is that the relations~(\ref{cov}) imply
\ba\label{pap}
\bar g_{ab}=\left(1+O(\phi^2)\right)g_{ab},\quad\quad\bar\phi=\left(1+O(\phi^2)\right)\phi, 
\ea
so that the on-shell asymptotic form of the new fields coincides with that of the original variables: $\bar\phi\rightarrow\phi$. Denoting the asymptotic form of the rescaled field as 
\ba
\bar\phi\sim\bar\alpha z^{\bar\Delta_-}+\bar\beta z^{\bar\Delta_+}, 
\ea
we thus have
\ba
\bar\alpha=\alpha\quad\quad\bar\beta=\beta,
\ea 
with the functional relation $\bar\beta(\bar\alpha)$ following from~(\ref{bc'}).

For the non-pathologic cases in which $Y>0$, and provided we restrict our study to potentials $\bar V$ which can be defined in terms of a potential $\bar P_-$ (see the discussion below~(\ref{pot})), one can argue that the stability of the nonminimally coupled system~(\ref{NM}) is controlled by the effective potential
\ba\label{Weff}
\tilde W(\phi)=U(\phi)-d\frac{Y''(0)}{2G}\phi^2-\frac{\bar\Delta_-}{2}\phi^2.
\ea
Eq.~(\ref{Weff}) is the generalization of~(\ref{W}) to the case of nonminimally coupled systems. The potential~(\ref{Weff}) determines the functional relation between $\bar\alpha$ and $\bar\beta$ very much like the potential~(\ref{W}) determines the relation between $\alpha$ and $\beta$ in a minimal system. In the definition of $\tilde W(\phi)$ we took into account the fact that $Y$ is given by~(\ref{bY}) near the conformal boundary.

In the presence of the quadratic deformations introduced in~(\ref{HG}),
\ba
Y&=&1-\xi G\phi^2\\\no
U&=&\frac{\bar\Delta_-}{2}\phi^2+\frac{\bar f}{2}\phi^2,
\ea
our stability condition for the nonminimally coupled AdS/CFT system coincides with the perturbative criteria, as it was for the minimally coupled case. In fact, the effective potential~(\ref{Weff}) is bounded below by $\tilde W\geq0$ provided 
\ba
\bar f+2d\xi\geq0
\ea
holds. The latter bound is equivalent to the perturbative requirement of absence of tachyons in the $AdS_{d+1}$ background, see Section~\ref{RES}, or to the requirement that there exists no IR Landau pole in the RG analysis, see~(\ref{fxi}) or~(\ref{xi}).

In summary, we have found that the system~(\ref{NM}) with the $AdS$ boundary conditions~(\ref{bc'})(\ref{as}) has a stable $AdS$ ground state if the potential~(\ref{Weff}) has a stable global minimum with $\tilde W\geq0$ (recall that $\tilde W(0)=0$ by definition). The validity of our proof is based on the existence of a stability theorem for minimally coupled scalar/gravity systems~\cite{des2}\cite{des3}\cite{des4}, as well as the assumptions that $Y>0$ and~(\ref{pap}). The former assumption is typically satisfied as long as our semiclassical approach is reliable, namely if $G\phi^2\ll1$. For the latter assumption to hold, it is crucial that the scalar $\phi$, or equivalently ${\cal O}$, has an unbroken flavor symmetry in the UV. In our case this is simply a discrete $Z_2$, but the result can be straightforwardly generalized to an arbitrary symmetry, and in particular to an arbitrary number of scalars.

\section{Summary of the results}

The main results of the present study can be summarized as follows:
\begin{itemize}
\item[i)] we developed a prescription that allows us to treat the $\Delta_\pm$ quantizations on a same footing. This prescription systematically encodes the CFT deformation $W$, see~(\ref{theory}), as a boundary term in the gravity dual. The precise mapping is given in~(\ref{result}) for $\Delta=\Delta_-$;
\item[ii)] for the particular case of doubletrace deformations we reproduced the RG flow found in field theory, and shown that nonminimal couplings to gravity are typically associated to a breaking of the conformal symmetry of the boundary theory, see~(\ref{a});
\item[iii)] a new class of excitations in the context of the gauge/gravity correspondence has been identified. These states have non-vanishing widths already at the leading $1/N$ order, and are mapped into IR non-normalizable modes (Gamow states) on the gravity side. When the dual theory flows between two fixed points of the RG, one such resonance naturally emerges as the remnant of the Nambu-Goldstone mode of dilatation invariance;
\item[iv)] we generalized an existing nonperturbative, classical stability condition for minimally coupled scalar/gravity systems on $AdS_{d+1}$ backgrounds to the case of nonminimally coupled theories: 
$AdS_{d+1}$ is stable if the effective potential~(\ref{Weff}) -- satisfying $\tilde W(0)=0$ by definition -- is such that $\tilde W\geq0$.
\end{itemize}

\paragraph{Note added}

After the first version of this work was submitted to the ArXiv, the authors of~\cite{stab} shown that the vacuum energy of the minimally coupled system defined in Section~\ref{stability} is controlled by the following effective potential:
\ba\label{EFFE}
V_{\rm{eff}}=a|\alpha|^{d/\Delta_-}+W(2\nu\alpha),
\ea
with $a>0$. This formula can be understood by recalling that multitrace deformations are mapped into deformations of the 1PI action, see Section~\ref{multi}: the effective potential of the deformed theory is the sum of the CFT potential for the classical field $2\nu\alpha$ (the first, bounded below, scale-invariant term in~(\ref{EFFE})) plus the explicit deformation (the second term in~(\ref{EFFE})). In the case of nonminimally coupled systems the function $W$ in~(\ref{EFFE}) is given by~(\ref{Weff}). The result~(\ref{EFFE}) clearly shows that a sufficient condition for the stability of $AdS_{d+1}$ is that the requirement iv) applies.

\acknowledgments
The author would like to thank T. Bhattacharya for interesting conversations and especially M. L. Graesser for numerous and helpful discussions. This work has been supported by the U.S. Department of Energy at Los Alamos National Laboratory under Contract No. DE-AC52-06NA25396.

\appendix

\section{Nonminimal couplings and surface terms}
\label{A}

The nonminimal Hawking-Gibbons term ensures that gravity admits a consistent variational problem, and in particular that no derivatives of the metric variation $\delta g_{ab}$ survive at the boundary $\partial M$ of the space-time manifold. 

Indeed, recall that the variation of the Ricci tensor contains a term of the form $\delta R= \nabla\nabla\delta g+\dots$ -- where the tensor structure has been neglected for simplicity and the dots stand for bulk terms -- that induces a boundary integral
\ba
\int_M\sqrt{g}\delta R= \int_{\partial M}\sqrt{h}n^a(\nabla\delta g)_a+\dots. 
\ea
The minimal Hawking-Gibbons integral $\int_{\partial M} \sqrt{h}K$ precisely cancels all the surface terms that include derivatives of the variation $\delta g$, and in particular ensures that no boundary terms survive if $\delta g|_{\partial M}=0$. 

In the presence of a nonminimal function $Y$ one finds a similar result. By varying the $YR$ term in~(\ref{NM}) and integrating by parts we have 
\ba
\delta(YR)&=&Y\nabla\nabla\delta g+\dots\\\no
&=&\nabla(Y\nabla\delta g)-\nabla Y\nabla\delta g+\dots\\\no
&=&\nabla(Y\nabla\delta g)+\nabla(\nabla Y\delta g)-(\nabla\nabla Y)\delta g+\dots.
\ea
The third term contributes to the equation of motion~(\ref{EOMM}) and can be discarded in our discussion, while the second vanishes if the metric is kept fixed at the boundary. Therefore, only the first term survives for $\delta g|_{\partial M}=0$, giving a surface integral
\ba
\int_{\partial M}\sqrt{h}n^a(Y\nabla\delta g)_a+\dots.
\ea
This is exactly of the same form as in the minimally coupled case, except for the appearance of an arbitrary factor $Y$. The variation of the nonminimal Hawking-Gibbons term in~(\ref{NM}) precisely cancels all the derivatives of the metric variation at the boundary $\partial M$.



\begin{thebibliography}{99}
 
 
 
 
\bibitem{first}
  O.~Aharony, M.~Berkooz and E.~Silverstein,
  JHEP {\bf 0108}, 006 (2001)
  [arXiv:hep-th/0105309].
O.~Aharony, M.~Berkooz and E.~Silverstein,
  Phys.\ Rev.\  D {\bf 65}, 106007 (2002)
  [arXiv:hep-th/0112178].

 

 \bibitem{Witten}
  E.~Witten,
  arXiv:hep-th/0112258.

 
\bibitem{preSS}
  M.~Berkooz, A.~Sever and A.~Shomer,
  JHEP {\bf 0205}, 034 (2002)
  [arXiv:hep-th/0112264].

 



  \bibitem{Mueck}
  W.~Mueck,
  Phys.\ Lett.\  B {\bf 531}, 301 (2002)
  [arXiv:hep-th/0201100].


 \bibitem{SS}
  A.~Sever and A.~Shomer,
  JHEP {\bf 0207}, 027 (2002)
  [arXiv:hep-th/0203168].

\bibitem{Maldacena}
  J.~M.~Maldacena,
  Adv.\ Theor.\ Math.\ Phys.\  {\bf 2}, 231 (1998)
  [Int.\ J.\ Theor.\ Phys.\  {\bf 38}, 1113 (1999)]
  [arXiv:hep-th/9711200].


 
 \bibitem{AdS/CFT}
  
 S.~S.~Gubser, I.~R.~Klebanov and A.~M.~Polyakov,
  Phys.\ Lett.\  B {\bf 428}, 105 (1998)
  [arXiv:hep-th/9802109].

 E.~Witten,
  Adv.\ Theor.\ Math.\ Phys.\  {\bf 2}, 253 (1998)
  [arXiv:hep-th/9802150].


\bibitem{Klebanov}
  I.~R.~Klebanov and E.~Witten,
  Nucl.\ Phys.\  B {\bf 556}, 89 (1999)
  [arXiv:hep-th/9905104].









\bibitem{PR}
  E.~Pomoni and L.~Rastelli,
  JHEP {\bf 0904}, 020 (2009)
  [arXiv:0805.2261 [hep-th]].
 
 










 
  \bibitem{cw}
  L.~Vecchi,
  Phys.\ Rev.\  {\bf D82}, 045013 (2010).
  [arXiv:1004.2063 [hep-th]].
  
  
  

 \bibitem{Kleba}
  A.~Dymarsky, I.~R.~Klebanov and R.~Roiban,
  JHEP {\bf 0508}, 011 (2005)
  [arXiv:hep-th/0505099].






  \bibitem{Petkou}
  G.~Arutyunov, S.~Penati, A.~C.~Petkou, A.~Santambrogio and E.~Sokatchev,
  Nucl.\ Phys.\  B {\bf 643}, 49 (2002)
  [arXiv:hep-th/0206020].


 
   \bibitem{Mack}
  G.~Mack,
  Commun.\ Math.\ Phys.\  {\bf 55}, 1 (1977).

   
 
 


 
  
  
 
 
 
 
 
 
 
 \bibitem{BF}
  P.~Breitenlohner and D.~Z.~Freedman,
  Annals Phys.\  {\bf 144}, 249 (1982).
  
  
  
  
  

  
 
 
 
\bibitem{GK}
  S.~S.~Gubser and I.~R.~Klebanov,
  Nucl.\ Phys.\  B {\bf 656}, 23 (2003)
  [arXiv:hep-th/0212138].
  

 
 
   \bibitem{Porrati}
  S.~Elitzur, A.~Giveon, M.~Porrati and E.~Rabinovici,
  JHEP {\bf 0602}, 006 (2006)
  [arXiv:hep-th/0511061].


  \bibitem{des2}
  T.~Hertog and S.~Hollands,
  Class.\ Quant.\ Grav.\  {\bf 22}, 5323 (2005)
  [arXiv:hep-th/0508181].

  
  
  \bibitem{des3}
  A.~J.~Amsel and D.~Marolf,
  Phys.\ Rev.\  D {\bf 74}, 064006 (2006)
  [Erratum-ibid.\  D {\bf 75}, 029901 (2007)]
  [arXiv:hep-th/0605101].
  
  
  
  \bibitem{des4}
  A.~J.~Amsel, T.~Hertog, S.~Hollands and D.~Marolf,
  Phys.\ Rev.\  D {\bf 75}, 084008 (2007)
  [Erratum-ibid.\  D {\bf 77}, 049903 (2008)]
  [arXiv:hep-th/0701038].

  
 
 
  
  \bibitem{HolRG}
  J.~de Boer, E.~P.~Verlinde and H.~L.~Verlinde,
  JHEP {\bf 0008}, 003 (2000)
  [arXiv:hep-th/9912012].

 
 
 \bibitem{Minces1}
  P.~Minces and V.~O.~Rivelles,
  Nucl.\ Phys.\  B {\bf 572}, 651 (2000)
  [arXiv:hep-th/9907079].


\bibitem{Minces2}
  P.~Minces and V.~O.~Rivelles,
  JHEP {\bf 0112}, 010 (2001)
  [arXiv:hep-th/0110189].

 
 
   \bibitem{Sundrum}
  A.~Lewandowski, M.~J.~May, R.~Sundrum,
  Phys.\ Rev.\  {\bf D67}, 024036 (2003).
  [hep-th/0209050].
  
  
    


    
  \bibitem{GM}
  S.~S.~Gubser and I.~Mitra,
  Phys.\ Rev.\  D {\bf 67}, 064018 (2003)
  [arXiv:hep-th/0210093].




   
   
 
 
 

\bibitem{Moroz}
  S.~Moroz,
  Phys.\ Rev.\  D {\bf 81}, 066002 (2010)
  [arXiv:0911.4060 [hep-th]].

 
 \bibitem{CL'}
 S.-J. Rey, talk at Strings 2007 conference (2007).
 
 
  \bibitem{CL}
  D.~B.~Kaplan, J.~W.~Lee, D.~T.~Son and M.~A.~Stephanov,
  Phys.\ Rev.\  D {\bf 80}, 125005 (2009)
  [arXiv:0905.4752 [hep-th]].


 
 
 
 \bibitem{mio}
  L.~Vecchi,
  Phys.\ Rev.\  D {\bf 78}, 085029 (2008)
  [arXiv:0712.1225 [hep-th]].

 
 
 
 \bibitem{Boucher}
  W.~Boucher,
  Nucl.\ Phys.\  B {\bf 242}, 282 (1984).

 
 
 
 

 
 
 
 
 
 
 \bibitem{Maeda}
  T.~Hertog and K.~Maeda,
  JHEP {\bf 0407}, 051 (2004)
  [arXiv:hep-th/0404261].

 
  
  \bibitem{des1}
  T.~Hertog and G.~T.~Horowitz,
  Phys.\ Rev.\ Lett.\  {\bf 94}, 221301 (2005)
  [arXiv:hep-th/0412169].

  
  
 
    
  
 

  
  
  \bibitem{Papa2}
  I.~Papadimitriou,
  JHEP {\bf 0705}, 075 (2007)
  [arXiv:hep-th/0703152].
  
  
  \bibitem{DeW}
  O.~DeWolfe, D.~Z.~Freedman, S.~S.~Gubser, A.~Karch,
  Phys.\ Rev.\  {\bf D62}, 046008 (2000).
  [hep-th/9909134].
  
 

 \bibitem{stab}
  T.~Faulkner, G.~T.~Horowitz, M.~M.~Roberts,
  Class.\ Quant.\ Grav.\  {\bf 27}, 205007 (2010).
  [arXiv:1006.2387 [hep-th]].
 
  
  
  
 \end{thebibliography}
 \end{document}